\newcommand{\vf}
{v_{\scriptscriptstyle F}}
\newcommand{\vfw}
{v_{{\scriptscriptstyle F},w}}
\newcommand{\vflocal}
{v_{{\scriptscriptstyle F},loc}}

\documentclass[prl,twocolumn,showpacs]{revtex4}

\usepackage{amsmath}
\usepackage[dvips,xdvi]{graphicx}
\usepackage{amssymb}
\usepackage{epsfig}

\catcode`\Š = \active \catcode`\š = \active \catcode`\Ÿ = \active
\catcode`\€ = \active \catcode`\… = \active \catcode`\† = \active
\catcode`\§ = \active \catcode`\Ž = \active \catcode`\ = \active
\catcode`\' = \active \catcode`\™ = \active \catcode`\ = \active
\catcode`\¿ = \active \catcode`\˜ = \active \catcode`\' = \active
\defŠ{\"a} \defš{\"o} \defŸ{\"u} \def€{\"A} \def…{\"O} \def†{\"U} \def§{\ss} \defŽ{\'{e}}
\def{\`{e}} \def'{\"{e}} \def™{\^{o}} \def{\^{e}} \def¿{\o} \def˜{\`{o}} \def'{\'{i}}

\begin{document}
\title{Critical velocity for superfluid flow across the BEC-BCS crossover}
\author{D. E. Miller, J. K. Chin, C. A.  Stan\footnote{Present address: Department of Chemistry and Chemical Biology,
 Harvard University, Cambridge, Massachusetts 02138}, Y. Liu, W. Setiawan, C. Sanner and W. Ketterle \footnote{Website: cua.mit.edu/ketterle\_group}}

\affiliation{Department of Physics, MIT-Harvard Center for
Ultracold Atoms, and Research Laboratory of Electronics, MIT,
Cambridge, MA 02139}

\date{\today}

\begin{abstract}

Critical velocities have been observed in an ultracold superfluid
Fermi gas throughout the BEC-BCS crossover. A pronounced peak of
the critical velocity at unitarity demonstrates that superfluidity
is most robust for resonant atomic interactions. Critical
velocities were determined from the abrupt onset of dissipation
when the velocity of a moving one dimensional optical lattice was varied. The
dependence of the critical velocity on lattice depth and on the
inhomogeneous density profile was studied.

\end{abstract}


\maketitle

The recent realization of the BEC-BCS crossover in ultracold
atomic gases \cite{Gior07} allows one to study how bosonic superfluidity
transforms into fermionic superfluidity. The critical velocity
for superfluid flow is determined
by the low-lying excitations of the superfluid.  For weakly bound
fermions, the (Landau) critical velocity is proportional to the
binding energy of the pairs, which increases monotonically along
the crossover into the BEC regime.  However, the speed of sound,
which sets the critical velocity for phonon excitations, is almost
constant in the BCS regime, but then decreases monotonically on
the BEC side, since the strongly bound molecules are weakly
interacting. At the BEC-BCS crossover, one expects a
rather narrow transition from a region where excitation of sound
limits superfluid flow, to a region where pair breaking
dominates. In this transition region,
the critical velocity is predicted to reach a maximum \cite{Sens06,Stri06,Stri07}.
This makes the critical velocity one of the few quantities
which show a pronounced peak across the BEC-BCS crossover in
contrast to the chemical potential, the transition temperature \cite{deMelo_PhysRevLett.71.3202},
the speed of sound \cite{zwer07,jose07} and the frequencies of shape
oscillations \cite{Grim07}, which all vary monotonically.

In this paper, we report the first study of critical velocities
across the BEC-BCS crossover, where a Feshbach resonance allows the magnetic
tuning of the atomic interactions,
and find that superfluid flow is
most robust near the resonance.  Our observation of a
pronounced maximum of the critical velocity is in agreement with
the predicted crossover between the two different mechanisms for
dissipation.

Critical velocities have been determined before in atomic BECs
perturbed by a stirring beam \cite{Rama99,Onof00,Rama01} as well
as by a 1D moving optical lattice \cite{ingu05}.  In both cases,
the inhomogeneous density of the harmonically trapped sample had
to be carefully accounted for in order to make quantitative
comparisons to theory.
Here we mitigate this problem by probing only the central region of our sample with
a tightly focused moving lattice formed from two intersecting laser beams.
For decreasing lattice depths, the
critical velocity increases and, at very small depths, approaches
a value which is in agreement with theoretical predictions.

In our experiments, we first create a superfluid of $^6$Li pairs
according to the procedure described in previous work
\cite{chin06}. Forced evaporative cooling of an even mixture of
the two lowest hyperfine states is performed at a magnetic field
of 822 G, on the BEC side of a broad Feshbach resonance centered
at $B_0 = 834$ G. This results in a nearly pure Bose-Einstein
condensate of $N\approx 5 \times 10^5$ pairs in a cross optical
dipole trap with harmonic trapping frequencies
$\nu_{x,y,z}\!=\!(65,45,50)$ Hz. The Fermi energy of the system is
$E_F\!=\!h \bar{\nu}\left( 6N\right)^{1/3} \!=\! h \times 7.6$ kHz.
To form the
moving lattice, we focus two phase-locked 1064 nm laser beams to
intersect at the sample with an angle of $\sim 90^\circ$ (see
Figure 1). The resulting 1D lattice has a spatial period of
$\lambda_L = 0.75\, \mu$m.
 A frequency difference between the two beams of
$\Delta\nu$ causes the lattice to move with velocity $v_L =
\lambda_L\,\Delta\nu$. The beams have $e^{-2}$ waists of 20 $\mu$m
and 60 $\mu$m respectively, and address a relatively homogeneous
region at the center of the cloud which has Thomas-Fermi radii
$R_{x,y,z}=(63,91,82)\, \mu$m.  The minimum density at the
position of the $e^{-2}$ waist is 42\% of the central density.

The lattice which necessarily varies in depth across the sample,
is characterized by its peak depth $V_0$ specified in units of
$E_F$ or the recoil energy $E_r = h^2/(8 m\lambda_L^2)= h \times 7.3$ kHz,
where $m$ is the molecular mass.
The lattice depth is calibrated using
Kapitza-Dirac scattering. Due to the inhomogeneity of the lattice,
the uncertainty is $40 \%$. The lattice depths explored in this
work are sufficiently small such that motion induced in the
laboratory frame is negligible, in contrast to \cite{mun07}.
\begin{figure}[]
\includegraphics[width=75mm]{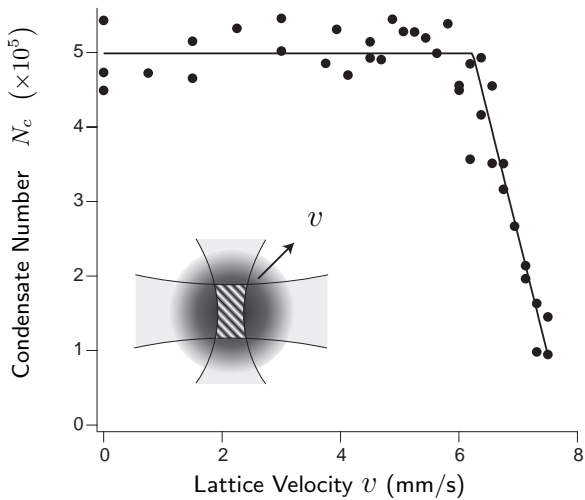}
\put(-117,98){\large$v$}
\put(-160,-3){\textsf{Lattice Velocity {\large$v$} (mm/s)}}
\put(-230,40){\textsf{\rotatebox{90}{{Condensate Number \rule{.1cm}{0cm} $N_c$ \rule{.1cm}{0cm}$\left(\times 10^5\right)$ }}}}
\caption{Onset of dissipation for superfluid fermions in a moving
optical lattice. (inset) Schematic of the experiment in which two
intersecting laser beams produced a moving optical lattice at the
center of an optically trapped cloud (trapping beams not shown).
Number of fermion pairs which remained in the condensate $N_c$
after being subjected
to a $V_0 = 0.2\, E_F$ deep optical lattice for 500 ms, moving
with velocity $v_L$, at a magnetic field of 822 G ($1/k_Fa =
0.15$)) An abrupt onset of dissipation occurred above a critical
velocity $v_c$, which we identify from a fit to Equation 1.}
\label{fig:dissipation}
\end{figure}

The lattice moving at a constant velocity is adiabatically ramped
up and held for a time $t$ up to 2 s, after which the lattice is
ramped down and all confinement is switched off. As in previous
work \cite{chin06}, a fast magnetic field ramp is used to reduce
strong interactions in order to probe the center-of-mass momentum
distribution of the pairs. Subsequently, absorption imaging is
done on the atomic resonance line at 730 G. A bimodal fit
reveals the number of pairs remaining in the condensate $N_c$, providing a 
measure of the heating incurred
during application of the moving lattice.

Figure \ref{fig:dissipation} illustrates the characteristic
dependence of dissipation on the velocity of the moving lattice.
At low velocities, the sample is unaffected. Above some critical
velocity $v_c$, dissipation sets in abruptly. We
determine $v_c$ from a fit of $N_c$
to the intersection of two
lines with slopes 0 and $\alpha$:
\begin{equation}
N_{cond} (v) = N_{cond}(0) \times \left[ 1 - max(0,\alpha \times
(v-vc))\right] \label{eq:Ncond}
\end{equation}
\noindent The critical velocity which we obtain from this
procedure is consistent for a large range of hold times, varying
by less than 15\% when the hold time $t$ is changed by a factor of
20.  We explore the BEC-BCS crossover by adiabatically ramping the magnetic field to
different values after evaporation and repeating the measurement as before. The
crossover is parameterized by the interaction parameter $1/k_Fa$, where $k_F$ is the
Fermi wavevector and $a$ is the B-field dependent s-wave scattering length \cite{bart05}.
Again, we observe a threshold for dissipation.

\begin{figure}[]
\includegraphics[width=80mm]{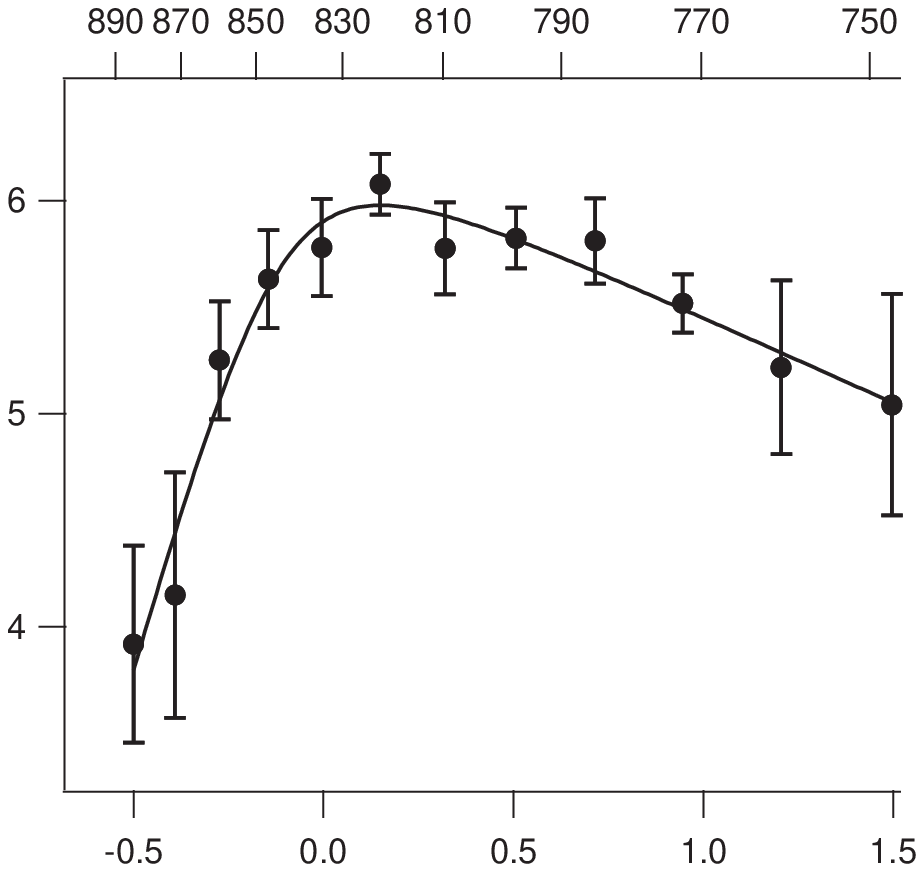}
\put(-175,-2){\textsf{Interaction Parameter $1/k_Fa$}}
\put(-160,205){\textsf{Magnetic Field (Gauss)}}
\put(-232,55){\rotatebox{90}{\textsf{Critical Velocity {\large$\;v_c$}  (mm/s)}}}
\caption{Critical velocities throughout the BEC-BCS crossover.
A pronounced maximum was found at resonance.  Data is shown for a
$V_0 = 0.2\, E_F$ deep lattice, held for t=500 ms. The solid line
is a guide to the eye.} \label{fig:bec-bcs}
\end{figure}

Figure \ref{fig:bec-bcs} shows the measured critical velocity
throughout the BEC-BCS crossover. The  maximum near resonance is
consistent with the picture of a crossover between two different
types of excitation, as discussed in the introduction, and confirms
that superfluidity is most robust on resonance.

\begin{figure}[]
\includegraphics[width=70mm]{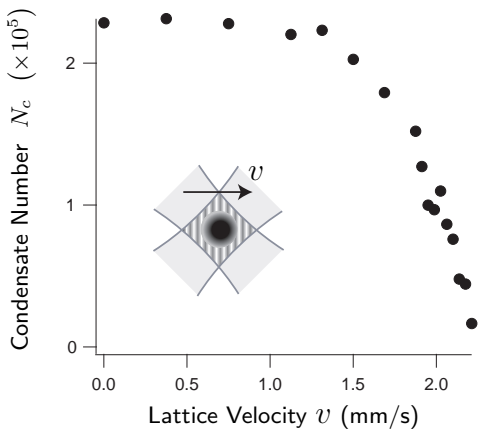}
\put(-140,-2){\textsf{Lattice Velocity {\large$v$} (mm/s)}}
\put(-193,25){\textsf{\rotatebox{90}{Condensate Number \rule{1mm}{0mm}$N_c$\rule{1mm}{0mm} $\left(\times 10^5\right)$ }}}
\put(-102,90){\large$v$}
\caption{Effects of density inhomogeneity on the critical velocity.  A configuration in which the lattice beams ($80 \mu$m) were larger than
the trapped sample ($37 \mu$m) results in loss in the condensate number $N_c$ at significantly lower velocity.
Data is shown for a $V_0 = 0.15\, E_F$ deep optical lattice held for 200 ms at a magnetic field of 822 G.
} \label{fig:inhomogeneous}
\end{figure}

To illuminate the role of the inhomogeneous density distribution,
we performed  experiments in which the entire sample was perturbed
by a uniform lattice. Lattice beams with $80\,\mu$m waists probed
a more tightly confined sample of $2\times10^5$ pairs, with
spatial extent R$_{TF}\simeq 37 \mu$m.
The onset of dissipation seen in Figure \ref{fig:inhomogeneous} is
still striking, but now loss is observed at much lower lattice
velocities, in spite of a larger Fermi energy $E_F\!=\! h\times 12.4\,$kHz.
Moreover, the onset of dissipation is slightly more gradual.  When
the magnetic field was varied across the Feshbach resonance, we
again found a maximum of the critical velocity near resonance. The
lowering of the critical velocity due to the inhomogeneous density
profile is expected, since at lower density, both the speed of
sound and (on the BCS side) the pairing energy decrease.  Although
the critical velocity should approach zero in the low density
wings of the cloud, we still observe a sudden onset of dissipation
at a finite velocity, similar to studies in Ref. \cite{Rama99,Onof00,Rama01}, where a
laser beam pierced through the whole condensate, but in contrast
to studies reported in \cite{ingu05}.

\begin{figure}[]
\includegraphics{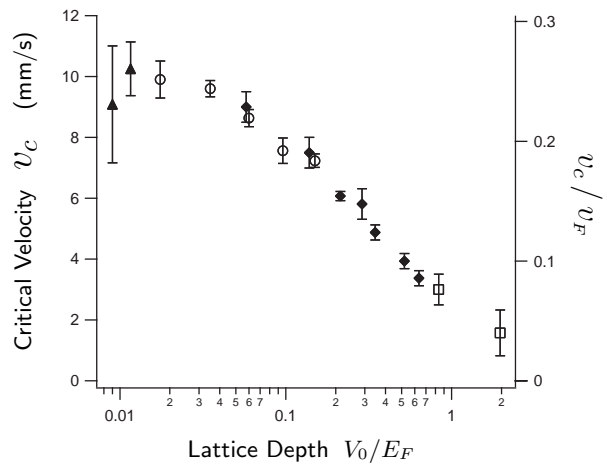}
\put(-226,45){\textsf{\rotatebox{90}{Critical Velocity {\Large$\;v_c\;\;$}  (mm/s)}}}
\put(-15,120){\rotatebox{-90}{\large${\displaystyle v_c\,/\,v_{\scriptscriptstyle F}}$}}
\put(-160,5){\textsf{Lattice Depth $\;V_0/E_F$}}
\caption{Critical velocities at different lattice depths. The
results show $v_c$ to be a decreasing function of lattice depth
$V_0$. In the limit of low $V_0$, $v_c$ converges to a maximum
value of 0.25 $\vf$. Data was taken near resonance, at 822 G
($1/k_Fa = 0.15$) for hold times t = 250 ms, 500 ms, 1000 ms, 2000
ms (squares, diamonds, circles, triangles). }
\label{fig:lattice_depths}
\end{figure}

In the limit of vanishing perturbation, the critical velocity
should be given by the Landau criterion.
In Figure \ref{fig:lattice_depths} we address
the effects of a finite lattice potential in the original lattice
configuration, as depicted in Figure \ref{fig:dissipation}.
The critical velocity is shown to be a decreasing function of $V_0$,
saturating in the limit of low lattice depth ($V_0 \leq 0.03\, E_F$).
This behavior is consistent with numerical simulation \cite{iane06,Stri07}.
Measurements at the smallest lattice depths had large
uncertainties, as the hold time required to observe a heating
effect of the lattice approached the natural lifetime of our
sample. For this reason  we studied the field dependence (Figure
\ref{fig:bec-bcs}) at an intermediate lattice depth, where $v_c$
was more well defined.

For comparison with theory we reference the local Fermi velocity
at the trap center $\vf \!=\! \vf^\prime \,(1+\beta)^{{\scriptscriptstyle -}1/4}$\linebreak $= 39$ mm/s,
where $\vf^\prime=\sqrt{2E_F/m}$ is the Fermi velocity of a non-interacting gas at the trap center,
and $\beta = -0.58$ is a universal parameter characterizing unitarity limited interactions \cite{Carl03,Astr04,Carl05}.
For vanishing lattice depth, the observed critical velocity at
unitarity approaches $v_c/\vf = 0.25$.
If we use the local Fermi velocity
$\vfw$  at the  $e^{-2}$ waist of the lattice,
we obtain $v_c/\vfw=0.34$ .
The difference between these values indicates the uncertainty due to residual density inhomogeneity.
The local speed of sound in a Fermi
gas at unitarity is
\begin{equation}
c_s = \vflocal\,(1+\beta)^{1/2}/\sqrt{3}=0.37\,\vflocal \;\;.
\end{equation}
The critical velocity for pair breaking is
\begin{equation}
v_{pair} = \left((\sqrt{\Delta^2+\mu^2}-\mu)/m\right)^{1/2} = 0.34\, \vflocal
\end{equation} with
$\Delta\!=\!0.50\,\vflocal^2/2m$ \cite{Carl03,Carl05} and
$\mu\!=\!(1+\beta)\,\vflocal^2/2m$. These two values should provide approximate
upper bounds to the critical velocity at unitarity \cite{Sens06,Stri06}.
It seems natural that the combination of both excitation
mechanisms lowers the critical velocity further.  Within these uncertainties, and
those of the density, the theoretical predictions agree with the experimental results.

Up until now, we have deferred a discussion of how the moving
lattice couples to the excitations.
In a pure system at zero
temperature, one would expect the excitation spectrum to exhibit discrete resonances,
where the perturbation couples only
to modes with the $k$-vector of the lattice.
On the other hand, at finite temperature, it is possible that the lattice drags
along thermal atoms which are point-like perturbations and can
create excitations at all $k$-vectors.
Our observation that the dissipation sets in at a certain threshold velocity and
increases monotonically with velocity
is consistent with the participation of the thermal component.

We further elucidated the role of thermal
excitations, by varying the temperature.
Gradually reducing the trap depth from $U_0$ to
$U_{min}$, during exposure to a lattice moving above $v_c$, will suppress the
accumulation of a thermal component. The lifetime
in this case exceeded that for a sample held at a fixed depth of
either $U_0$ or $U_{min}$.  For Bose-Einstein condensates,
theoretical papers emphasized the role of the thermal component in
the Landau damping process in a moving lattice
\cite{griff04,niku05}.  This was confirmed qualitatively in an
experiment at Florence \cite{ingu05} in which the lifetime
of the sample was drastically improved by eliminating the thermal atoms.

\begin{figure}[]
\includegraphics[width=70mm]{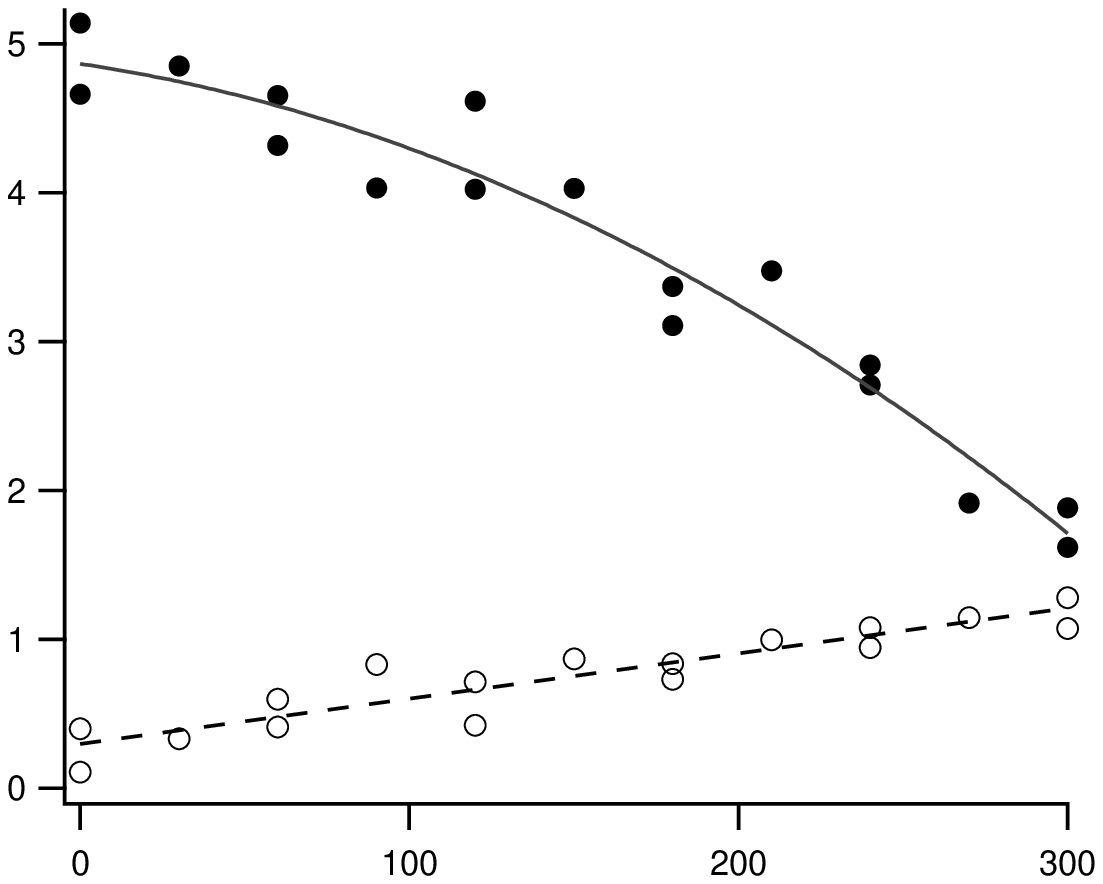}
\put(-208,50){\rotatebox{90}{Atom Number  ($\times 10^5$)}}
\put(-130,-3){Hold Time (ms)}
\caption{Number of pairs which remained in the condensate $N_c$
(filled circles) and thermal component $N_{th}$ (open circles)
after being held in a $V_0 = 0.35\, E_F$ deep optical lattice
moving with velocity $v_L=6$ mm/s for a variable hold time. The
thermal component shows a linear increase (dashed line), whereas
$N_{cond}$ showed an accelerated loss, and is fit to a quadratic function (solid line).
\label{fig:timedependence}
}
\end{figure}

In our experiments, the clouds heated up during the exposure to
the moving lattice. Figure \ref{fig:timedependence} shows the
increase in the number of thermal atoms and the loss in the total
number of atoms due to evaporative cooling.  In an idealized
model, where density is fixed,
constant dissipation would result in a linear
decrease in the number of atoms due to evaporative cooling.  Our
data show an accelerated decrease, possibly reflecting increased
dissipation due to the increasing fraction of thermal atoms.
However, an accurate model should include the change in density
(and therefore critical velocity) during the exposure time.
Additional impurity atoms (e.g. sodium atoms) could cause
dissipation even at zero temperature and would allow more
controlled studies of the dissipation mechanism.  Unpaired
atoms in clouds with population imbalance may not play this role
because of  phase separation effects \cite{shin06}.

Another possible dissipation mechanism in a lattice is the creation of two excitations
through a dynamical or modulational instability.  Such an
instability \cite{Wu_PhysRevA.64.061603} occurs already for weakly
interacting particles moving
through a lattice with momentum $q$, when they collide and scatter
into states with  momenta $q \pm \delta q$, analogous to optical
parametric generation \cite{camp06}.  This process is
energetically possible only
above $0.5\, q_B$, where $q_B = h/2\lambda_L$ is the Bragg
momentum which defines the edge of the Brillouin zone. This
corresponds to a velocity $v = 11$ mm/s for fermion pairs (and
twice this value for single atoms). Since the highest critical velocities
we observe are slightly below this
threshold,  and strongly decrease already for relatively small
$V_0$, it is very unlikely, that dynamical instabilities
play a role in our experiments.  Moreover, such instabilities should
be strongly modified by Pauli blocking. For our ratio of local Fermi
momentum to the Bragg momentum of 0.9, the first band is nearly full in the center of the cloud.
For Bose-Einstein
condensates, it has been recently predicted \cite{Altm05} and
experimentally shown  \cite{mun07} that strong interactions can
lower the threshold for the dynamical instability, close to the
Mott-insulator transition.  The range of 1-D lattice depths explored
here ($V_0 \leq 2\,E_r$) is far from the 1D Mott-insulator regime.
We have observed the loss of coherence which typically
accompanies the superfluid to Mott insulation transition to occur
only beyond $V_0 \simeq 25\,E_r$.

In conclusion, we have used a novel optical lattice geometry to
determine critical velocities in the BEC-BCS crossover without the
complications of strong density inhomogeneity.  This configuration could be
applied to studies in atomic Bose gases which so far have been
limited by the inhomogeneous density
\cite{Rama99,Onof00,Rama01,ingu05}. In addition, it would be
interesting to study dynamical instabilities for fermions and
the role of Pauli blocking.  The authors would like to
thank Aviv Keshet for experimental assistance. This research has
been supported by the NSF and the Office of Naval Research.

\bibliographystyle{prsty}

\bibliography{refs}

\begin{thebibliography}{10}

\bibitem{Gior07}
S. Giorgini, L.~P. Pitaevskii, and S. Stringari, preprint condmat/0706.3360  .

\bibitem{Sens06}
R. Sensarma, M. Randeria, and T.-L. Ho, Phys. Rev. Lett. {\bf 96},  090403
  (2006).

\bibitem{Stri06}
R. Combescot, M.~Y. Kagan, and S. Stringari, Phys. Rev. A {\bf 74},  042717
  (2006).

\bibitem{Stri07}
A. Spuntarelli, P. Pieri, and G.~C. Strinati, pre-print /condmat/0705.2658  .

\bibitem{deMelo_PhysRevLett.71.3202}
C.~A.~R. S\'a~de Melo, M. Randeria, and J.~R. Engelbrecht, Phys. Rev. Lett.
  {\bf 71},  3202  (1993).

\bibitem{zwer07}
R. Haussmann, W. Rantner, S. Cerrito, and W. Zwerger, Phys. Rev. A {\bf 75},
  023610  (2007).

\bibitem{jose07}
J. Joseph {\it et~al.}, Phys. Rev. Lett. {\bf 98},  170401  (2007).

\bibitem{Grim07}
A. Altmeyer {\it et~al.}, Phys. Rev. Lett. 98, 040401 (2007) {\bf 98},  040401
  (2007).

\bibitem{Rama99}
C. Raman {\it et~al.}, Phys. Rev. Lett. {\bf 83},  2502  (1999).

\bibitem{Onof00}
R. Onofrio {\it et~al.}, Phys. Rev. Lett. {\bf 85},  2228  (2000).

\bibitem{Rama01}
C. Raman {\it et~al.}, J. Low Temp. Phys. {\bf 122},  99  (2001).

\bibitem{ingu05}
L. De~Sarlo {\it et~al.}, Phys. Rev. A {\bf 72},  013603  (2005).

\bibitem{chin06}
J. Chin {\it et~al.}, Nature {\bf 443},  961  (2006).

\bibitem{mun07}
J. Mun {\it et~al.}, preprint condmat/0706.3946  .

\bibitem{bart05}
M. Bartenstein {\it et~al.}, Phys. Rev. Lett. {\bf 94},  103201  (2005).

\bibitem{iane06}
S. Ianeselli, C. Menotti, and A. Smerzi, J. Phys. B {\bf 39},  S135  (2006).

\bibitem{Carl03}
J. Carlson, S.-Y. Chang, V.~R. Pandharipande, and K.~E. Schmidt, Phys. Rev.
  Lett. {\bf 91},  050401  (2003).

\bibitem{Astr04}
G.~E. Astrakharchik, J. Boronat, J. Casulleras, and S. Giorgini, Phys. Rev.
  Lett. {\bf 93},  200404  (2004).

\bibitem{Carl05}
J. Carlson and S. Reddy, Phys. Rev. Lett. {\bf 95},  060401  (2005).

\bibitem{griff04}
S. Tsuchiya and A. Griffin, Phys. Rev. A {\bf 70},  023611  (2004).

\bibitem{niku05}
S. Konabe and T. Nikuni, J. Phys. B {\bf 39},  S101  (2005).

\bibitem{shin06}
Y. Shin {\it et~al.}, Phys. Rev. Lett. {\bf 97},  030401  (2006).

\bibitem{Wu_PhysRevA.64.061603}
B. Wu and Q. Niu, Phys. Rev. A {\bf 64},  061603  (2001).

\bibitem{camp06}
G.~K. Campbell {\it et~al.}, Phys. Rev. Lett. {\bf 96},  020406  (2006).

\bibitem{Altm05}
E. Altman {\it et~al.}, Phys. Rev. Lett. {\bf 95},  020402  (2005).

\end{thebibliography}

\end{document}